\documentclass[%
 aip,
 amsmath,amssymb,
 reprint,%
]{revtex4-1}

\usepackage{graphicx}
\usepackage{dcolumn}
\usepackage{bm}
\usepackage{braket}

\usepackage[utf8]{inputenc}
\usepackage[T1]{fontenc}
\usepackage{mathptmx}

\begin{document}

\preprint{AIP/123-QED}

\title[A novel CMB setup for strong-field-seeking molecules]{A novel crossed-molecular-beam experiment for investigating reactions of state- and conformationally selected strong-field-seeking molecules\vspace{+0.3cm}}

\thanks{Dedicated to Prof. J\"urgen Troe on the occasion of his 80$^\mathrm{th}$ birthday.}

\author{L. Ploenes}

\author{P. Stra\v{n}\'ak}%

\affiliation{Department of Chemistry, University of Basel, Klingelbergstrasse 80, 4056 Basel, Switzerland\looseness=-1
}%

\author{H. Gao}%
\affiliation{Department of Chemistry, University of Basel, Klingelbergstrasse 80, 4056 Basel, Switzerland\looseness=-1 
}%
\affiliation{Present address: Beijing National Laboratory for Molecular Sciences (BNLMS), Institute of Chemistry, Chinese Academy of Sciences, Beijing 100190, People’s Republic of China\looseness=-1
}%

\author{J. K\"upper}%
\affiliation{Center for Free-Electron Laser Science, Deutsches Elektronen-Synchrotron DESY, Notkestra\ss e 85, 22607 Hamburg, Germany\looseness=-1
}%
\affiliation{Center for Ultrafast Imaging, Universit\"at Hamburg, Luruper Chaussee 149, 22761 Hamburg, Germany\looseness=-1
}%

\affiliation{Department of Physics, Universit\"at Hamburg, Luruper Chaussee 149, 22761 Hamburg, Germany\looseness=-1
}%
\affiliation{Department of Chemistry, Universit\"at Hamburg, Martin-Luther-King-Platz 6, 20146 Hamburg, Germany\looseness=-1 \vspace{0.7cm}}

\author{S. Willitsch}
 \homepage{Electronic  address: stefan.willitsch@unibas.ch}
\affiliation{%
Department of Chemistry, University of Basel, Klingelbergstrasse 80, 4056 Basel, Switzerland\looseness=-1 
}%

\date{\today}

\begin{abstract}
The structure and quantum state of the reactants have a profound impact on the kinetics and dynamics of chemical reactions. Over the past years, significant advances have been made in the control and manipulation of molecules with external electric and magnetic fields in molecular-beam experiments for investigations of their state-, structure- and energy-specific chemical reactivity. Whereas studies for neutrals have so far mainly focused on weak-field-seeking species, we report here progress towards investigating reactions of strong-field-seeking molecules by introducing a novel crossed-molecular-beam experiment featuring an electrostatic deflector. The new setup enables the characterisation of state- and geometry-specific effects in reactions under single-collision conditions. As a proof of principle, we present results on the chemi-ionisation reaction of metastable neon atoms with rotationally state-selected carbonyl sulfide (OCS) molecules and show that the branching ratio between the Penning and dissociative ionisation pathways strongly depends on the initial rotational state of OCS.
\end{abstract}

\maketitle

\section{\label{sec:level1}Introduction \protect}
The kinetics and dynamics of a chemical reaction strongly depends on the properties of the reactant molecules including their collision energy, internal quantum states, geometries as well as alignment and orientation. In the last decades, significant progress has been achieved in unravelling the details of bimolecular reactions under single-collision condition,\cite{levine05a,brouard12a} notably by the development and continuous improvement of crossed-molecular-beam (CMB) experiments.\cite{bull55a, taylor55a,lee87a,casavecchia99a,ashfold06a,liu06a,huilin17a,wang18a,li20a}  
The information to be gained from such studies substantially depends on the ability to control and prepare the reactants. While in conventional molecular-beam setups the molecules are internally cooled to rotational and translational temperatures of typically a few Kelvin in supersonic gas expansions, they usually still populate a range of different quantum states and, in the case of complex species, even different molecular conformations. Reaction cross sections obtained in this way are thus averages over a range of quantum states and molecular configurations obscuring the precise influence of these degrees of freedom on chemical reactivity. \\
Significant progress in the control of the reactant molecules was recently achieved by utilizing external electric or magnetic fields in molecular-beam experiments. Selection of individual rotational states has been achieved using electrostatic-multipole focusers.\cite{gandhi86a, Reuss88a,  parker89a,vanLeuken95a,vanLeuken96a,gijsbertsen05a,tsai10a,brouard14a} The implementation of Stark\cite{bethlem99a, gilijamse06a,Scharfenberg10a} and Zeeman decelerators\cite{vanhaecke07a, narevicius07a,wiederkehr11a,dulitz14,cremers17a} in CMB setups enabled the measurement of state-to-state scattering cross sections with unprecedented collision-energy resolution.\cite{kirste12a,vonZastrow14a,vogels18a,deJongh20a,plomp20a} Moreover, the development of merged-beam experiments lead to an increased understanding of chemi-ionisation reactions of excited rare gas atoms with state-selected and even oriented atoms and molecules at very low collision energies.\cite{henson12a,bertsche14a,klein17a,gordon18a, margulis20a,zou20a,gordon20a, paliwal21a} \\
All of these studies exploited the focusing effects obtained by the interaction of an inhomogeneous external field with the electric or magnetic dipole moment of the molecules in weak-field seeking quantum states, i.e., states whose energy increases with increasing field strength, and focused on di- or small polyatomic systems. Of additional interest in chemistry are also larger polyatomic molecules featuring different stereoisomers. Due to the coupling between their closely spaced rotational levels, these molecules are usually exclusively strong-field-seeking at experimentally relevant field strengths. Such strong-field-seekers are not amenable to the techniques described above because the laws of electrodynamics do not allow a maximum for static fields in free space and hence no transverse stability along the beam axis can be achieved.\\
For the manipulation of strong-field-seeking molecules in a beam, alternating-gradient focusers were developed which enabled the successful control of diatomic and small polyatomic molecules such as CO, YbF, CaF, NH$_3$ and OH, \cite{kakati71a, bethlem02a, tarbutt04a, bethlem06a, wohlfart08a, wall09a} but also allowed for state selection and deceleration of larger species like benzonitrile\cite{wohlfart08b,putzke11a} and even spatial separation of conformers of molecules like 3-aminophenol.\cite{filsinger08a,filsinger11a} In addition, microwave lens systems were implemented for the selection of strong-field-seeking states.\cite{odashima10a,merz12a,spieler13a,enomoto19a} Moreover, laser-based optical techniques were applied for the deflection and deceleration of neutral, strong-field-seeking molecules.\cite{stapelfeldt97a, zhao00a, fulton04a, sun15a}
Furthermore, electrostatic deflection of strong-field-seeking molecules in inhomogeneous electric fields\cite{chang15a} enabled not only the selection of individual rotational states in small systems,\cite{nielsen11a,horke14a,trippel15a} but also the spatial separation of individual conformers of more complex species like different aromatic compounds,\cite{filsinger08a,filsinger09a,trippel18a} 2,3-dibromobuta-1,3-diene,\cite{kilaj20a} methyl vinyl ketone\cite{wang20} and dipeptides.\cite{teschmit18a} \\ 
Recently, we successfully applied this method in our laboratory\cite{willitsch17a} to the study of reactive collisions of conformationally selected 3-aminophenol molecules with trapped Ca\textsuperscript{+} ions\cite{chang13a, roesch14a} and different nuclear-spin isomers of water with N\textsubscript{2}H\textsuperscript{+} ions.\cite{kilaj18a}  We are now extending this approach to the investigation of reactions in which all reactants are neutral and report the development of a novel CMB setup featuring an electrostatic deflector. This new experiment enables the investigation of the dynamics and kinetics of a wide variety of reactions of state- and conformer-selected strong-field-seeking species with neutral reaction partners and allows for the characterisation of stereochemical effects in a single-collision environment. We demonstrate the capabilities of the new setup with a proof-of-principle experiment on the chemi-ionisation reaction of metastable neon atoms with rotational-state-selected carbonyl sulfide (OCS) molecules and show that the branching ratio between the Penning- and dissociative-ionisation reaction pathways strongly depends on the initial rotational state of OCS.

\section{\label{sec:level2}Experimental Setup \protect}
\begin{figure}[h]
\centering
\includegraphics[width=\columnwidth]{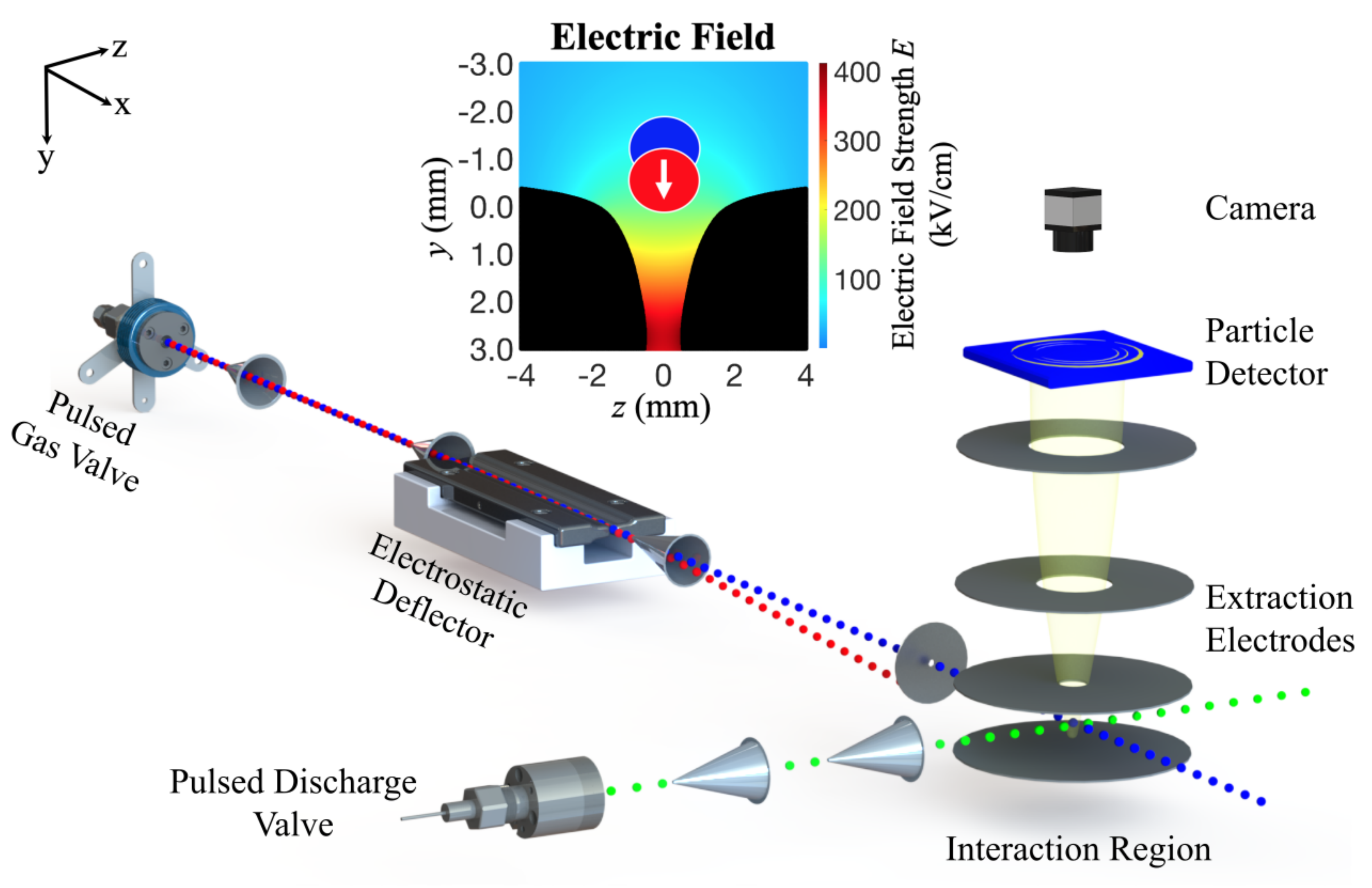}
\caption{Schematic of the novel crossed-molecular-beam setup for investigating reactions of strong-field-seeking molecules. An electrostatic deflector enables the spatial separation of individual quantum states or conformers of molecules, schematically indicated by the red and blue dots, in a pulsed molecular beam. In an interaction region, the spatially separated molecules are overlapped with a second molecular beam containing metastable or radical co-reactants (green dots) produced by a pulsed discharge valve. Reaction products are detected by time-of-flight mass spectrometry and velocity-mapped ion imaging. The inset shows a cross section of the inhomogeneous electric field in the deflector and the approximate position of the molecular beam separating into different components.}
\label{fig:Figure1}
\end{figure}

A schematic of the novel CMB apparatus is depicted in Fig.~\ref{fig:Figure1} and a construction drawing is shown in Fig.~\ref{fig:Figure2}. The setup features two molecular beams which intersect at an angle of 90$^\circ$ in the centre of a velocity-map-imaging\cite{eppink97a} (VMI) time-of-flight mass spectrometer (TOF-MS). The distinctive feature of the new setup is an electrostatic deflector integrated into one of the molecular beams which enables the spatial separation of different rotational states or conformers prior to reaction.\cite{chang15a}  For the generation of radicals and metastable-rare-gas reactants, the second molecular-beam source is equipped with a home-built discharge valve.

\subsection{\label{sec:level3}Spatial Separation of Strong-Field-Seekers: The Electrostatic Deflector}
\begin{figure*}[!ht]
\centering
\includegraphics[width=\textwidth]{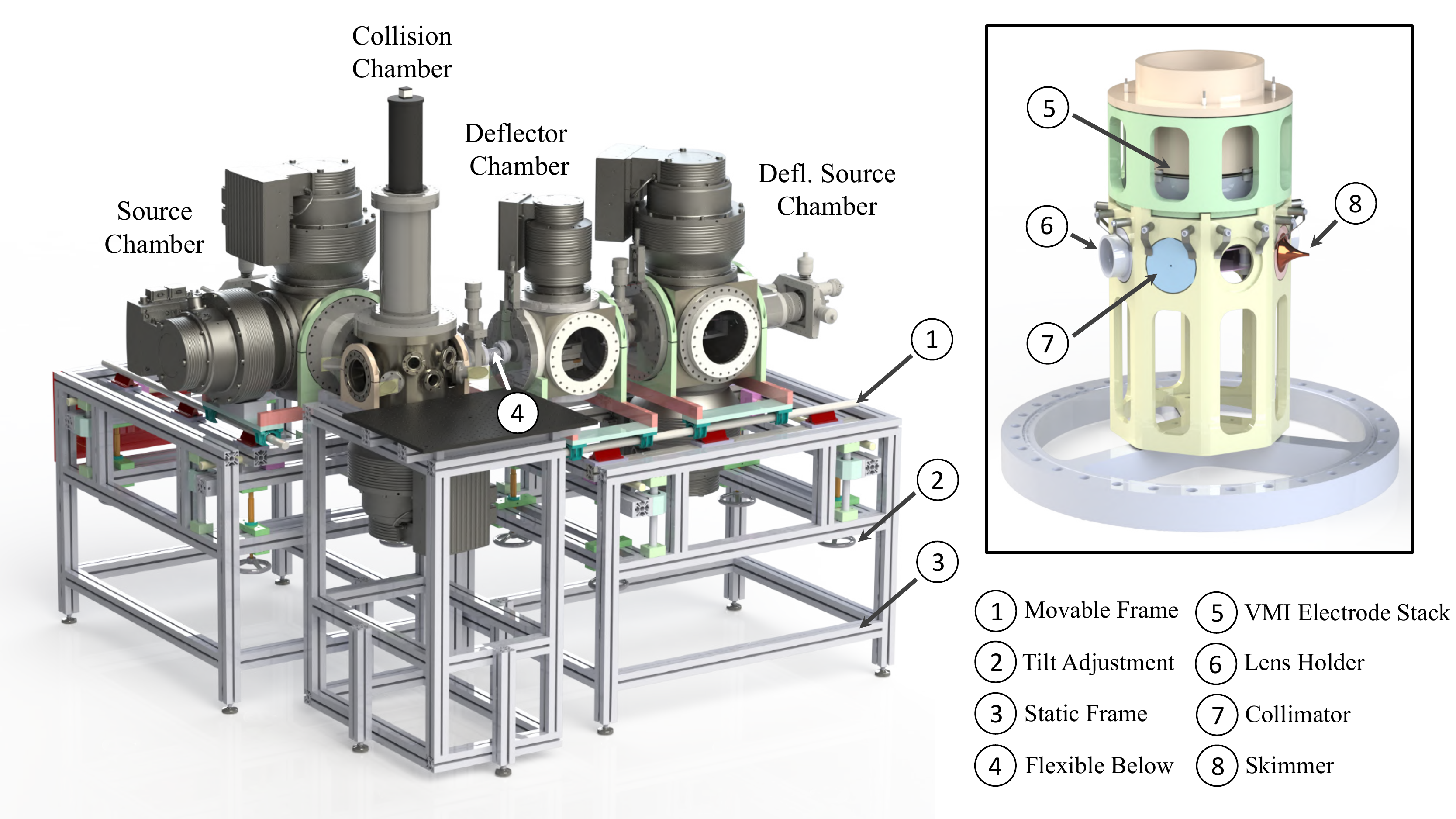}
\caption{Construction drawing of the CMB setup showing the vacuum assembly. The assembly for the deflected molecular beam is mounted on a movable frame which can be tilted with an adjustment wheel. Thus, different parts of the deflected beam containing different compositions of conformers or rotational states can be overlapped with the second molecular beam in the collision region. The inset shows the centre of the collision chamber housing an octagonal support structure which holds the ion-extraction electrodes and allows for mounting of skimmers, collimators and optical lenses.}
\label{fig:Figure2} 
\end{figure*}

The electrostatic deflector used in the present experiment is based on the design described in Ref. \onlinecite{kienitz17a}. Briefly, it consists of two parallel electrodes oriented along the molecular-beam axis with a total length of 15~cm spaced 1.5~mm apart (Fig.~\ref{fig:Figure1}).
Electric-potential differences of up to $\Delta V=$~40~kV can be applied across the electrodes creating a strongly inhomogeneous electric field with strengths up to $E=$~350~kV/cm (inset in Fig.~\ref{fig:Figure1}). A molecular beam is generated by a supersonic expansion from an Even-Lavie pulsed gas valve\cite{even00a} in a source chamber. After passing a skimmer with an orifice diameter of 3~mm separating the source and deflector chambers, the molecular beam is further constrained by two conical skimmers with orifice diameters of 1.5~mm mounted before and after the deflector. \\
Molecules passing through the deflector experience a force perpendicular to their propagation direction (the $y$-axis in Fig.~\ref{fig:Figure1}). The extent of deflection for a specific applied electric field depends on the ratio of the effective, space-fixed, dipole moment $\mu_{\text{eff}}$ of the molecule to its mass $m$.\cite{chang15a} In the present experiment, weak- and strong-field-seeking molecules are deflected into the positive and negative $y$~directions, respectively. For species with different effective dipole moments, e.g., molecules in different rotational states or different conformational isomers of the same molecule averaged over many rotational states, a spatial separation along the transverse direction of the molecular beam is thus achieved,\cite{chang15a} which further increases during free flight over a distance of 45~cm between the end of the deflector and the reaction region. Along this way, the molecular beam passes through another collimator consisting of an aperture with a diameter of 1.5~mm placed 8~cm before the centre of the collision region (see inset in Fig.~\ref{fig:Figure2}).  \\
The source and deflector chambers are mounted on a movable frame which can be tilted with respect to the collision chamber (see Fig.~\ref{fig:Figure2}). The position of the frame can be read out with an accuracy of 10~$\mu$m using a digital micrometer placed near the tilt adjustment indicated in Fig.~\ref{fig:Figure2}. By tilting the deflector assembly, different parts of the deflected molecular beam which contain molecules in different rotational states or different conformers can be overlapped with the second beam in the collision chamber so that state- or conformer-selected reactions can be studied.

\subsection{\label{sec:level4}Metastable-Atom and Radical Source}
The second beam containing the reaction partner of the deflected molecular species is generated by a solenoid-based pulsed valve.\cite{grzesiak18a} For the  production  of  internally  cold and  dense  beams  of  radicals  or  metastable  rare  gas  atoms, this valve was combined with either a dielectric barrier discharge (DBD) or a plate discharge\cite{stranak21a} based on the design of Ref. \onlinecite{ploenes16a}. The valve can operate at repetition rates up to 200~Hz and backing pressures $>$50~bar. The discharge is stabilised by an electron-ejecting filament. Unwanted ions in the plasma are deflected by a bias voltage of 500 V applied to the skimmer in the source chamber. 

\subsection{\label{sec:level5}Collision Chamber and Detection System}
The collision chamber houses a central octagonal support structure (inset in Fig.~\ref{fig:Figure2}) serving as the fixation and alignment point for both molecular beams, for all laser beams inserted into the experiment and for the extraction electrodes of the ion-detection system. Each side of the octagon features clip mounts to which skimmers, collimators or optical lenses can be attached reproducibly without losing the alignment of the particle and laser beams. This renders the setup flexible for adjusting experimental parameters such as the size of the collision and laser-ionisation volumes which have a strong effect on signal levels, detector resolution and selection of different parts of the deflected molecular beam.  \\
TOF mass spectrometry is employed for the characterisation of the reaction products. Additionally, information about the dynamics of reactions can be gained using VMI. After ionisation of the reaction products, either by using a suitable laser or directly in chemi-ionisation reactions, the ions are accelerated by an electrostatic lens towards a position-sensitive detector consisting of a 75~mm diameter multichannel-plate (MCP) stack coupled to a phosphor (P46) screen (Photek). The electrostatic lens system is based on the design described in Ref. \onlinecite{townsend03}. The VMI system was calibrated using the dissociation of molecular oxygen following Rydberg excitation.\cite{eppink97a} The extraction fields can be switched within a few nanoseconds using two independent high-voltage switches (Behlke).
In TOF-MS mode, the signal from the MCPs is enhanced by a preamplifier (x10) before being averaged by a fast oscilloscope. In VMI mode, the phosphor screen is read out by a fast camera (IDS UI-3040CP-M-GL Rev.2). The resulting image is thresholded, centroided and event-counted by a home-written software allowing real-time image analysis up to an experimental repetition rate of 1000~Hz for image sizes up to 1024x1024 pixels.

\subsection{\label{sec:level6} Operation at High Repetition Rates}
A prerequisite for achieving a large spatial separation of the deflected molecules is the long free-flight distance between the deflector and the reaction region of 45~cm in the present setup. In total, the distance between the molecular-beam source and the detection region sums up to $\approx$1~m, which is significantly longer than in conventional CMB setups.\cite{casavecchia99a,ashfold06a,liu06a,huilin17a,wang18a,li20a} This results in low beam densities in the interaction region and hence a reduced number of collision events. Therefore, the present setup has been designed to operate at higher repetition rates than the typical 10~Hz used in many previous setups in order to compensate for the inherent loss in sensitivity by faster data accumulation.  \\  
Both molecular beam sources can operate at frequencies up to 200 Hz. To ensure low enough pressures for efficient supersonic cooling of the molecules in the source region in spite of the high duty cycle, each source chamber features two magnetically levitated turbo pumps (Oerlikon Leybold MAG W~2200) with a pumping speed of 2200~l/s each. The collision and deflector chambers are differentially pumped with 2200~l/s and a 600~l/s  magnetically levitated turbo pumps, respectively (Oerlikon Leybold MAG W~2200 and MAG W~600). All turbo pumps are backed up by a single oil-free and corrosion resistant double-stage root pump (Pfeiffer A604) with a total pumping speed of 540~m\textsuperscript{3}/h. All turbo pumps and the backing pump are purged with dry N\textsubscript{2} to prevent corrosion. At a repetition rate of 200 Hz, the pressure of both source chambers is typically maintained in the 10\textsuperscript{-5}~mbar range and the pressure of the collision chamber does not exceed 5x10\textsuperscript{-7}~mbar which  is adequate for crossed-molecular-beam studies. \\ 
Multiphoton laser ionisation of molecules in the collision chamber is performed using a fs laser (Clark MXR, pulse duration 150~fs, wavelength 775~nm, repetition rates up to 1~kHz).

\subsection{\label{sec:level7} Monte-Carlo Trajectory Simulations of Deflected Molecules}
For the analysis of the experimental results, trajectories of molecules travelling from the source through the electrostatic deflector up to the reaction region were simulated using a Monte-Carlo approach implemented in home-written code.\cite{filsinger09b, chang13a, roesch14a, chang15a, kilaj20a} \\
Briefly, the Stark energy of the individual rotational states of the molecules in a specific external electric field was computed using the CMI Stark software package.\cite{chang14a} The electrostatic field of the deflector was calculated using COMSOL Multiphysics. The force $F$ acting on the molecules was derived from the relation $\vec{F} =-\vec{\nabla} W_\text{Stark}(E)=\mu_{\text{eff}}\cdot\vec{\nabla}E$, where $W_\text{Stark}(E)$ is the Stark energy in the electric field $E$, and the trajectories were calculated by numerical integration of the equations of motion using a Runge–Kutta algorithm. The initial velocity of the molecular beam was measured for each experiment using the calibrated VMI system. The dipole moment and rotational constant of OCS used in the simulations were taken from Ref. \onlinecite{reinartz74} and Ref. \onlinecite{larsen74}, respectively. \\
Trajectory simulations were performed for individual quantum states yielding state-resolved density profiles of the molecular beam in the detection region. Simulations of molecules in individual rotational states were performed fully parallelised to speed up effective computation times. Thermally averaged deflection profiles were generated by weighting the state-resolved density profiles according to the thermal state populations at a given rotational temperature. The rotational temperature of the molecules was determined by matching simulated thermally averaged density profiles to the experimental data by a least-square fitting procedure.
\\

\section{\label{sec:level8}Results and Discussion \protect}

\subsection{\label{sec:level9}State-specific Spatial Separation of OCS Molecules}
\begin{figure}[!h]
 \centering
 \includegraphics[width=\columnwidth]{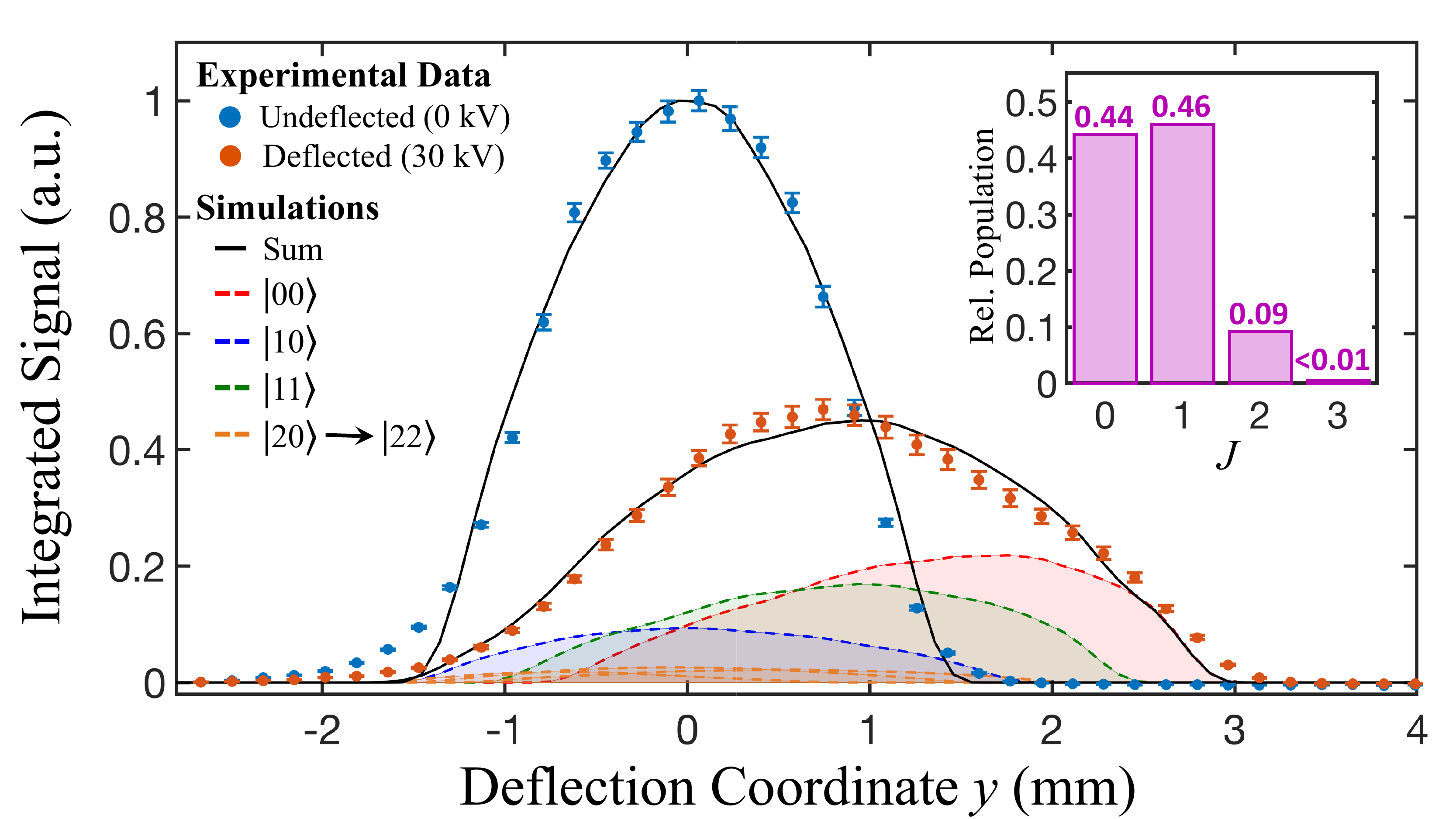}
 \caption{Experimental deflection profiles of OCS at deflector voltages 0~kV and 30~kV (blue and red data points, respectively) with corresponding Monte-Carlo trajectory simulations for different rotational states $\ket{JM}$ (dashed lines). The black lines represent the sum of the individual rotational-state-specific deflection profiles weighted by the rotational populations assuming a rotational temperature of T$_\text{rot}=0.55$~K. The error bars represent the standard error of ten individual measurements in which each data point was averaged over 1000 experimental cycles. Inset: Thermal populations of rotational levels of OCS with quantum number $J$ at T$_\text{rot}=0.55$~K.}
 \label{fig:Figure3} 
 \end{figure}

A pulsed molecular beam of 2000~ppm OCS (Sigma Aldrich) seeded in helium was produced by expanding 10~bar of the gas mixture into vacuum. The resulting supersonic jet of OCS molecules was guided through the electrostatic deflector and then multiphoton-ionised with fs-laser radiation. The resulting ions were detected by TOF mass spectrometry. Integrating the TOF peak of the OCS\textsuperscript{+} product for different tilting angles of the molecular beam machine yielded the beam-density profiles shown in Fig.~\ref{fig:Figure3}. The deflection coordinate $y$ represents the offset from the center of the undeflected molecular beam. Experimental beam-density profiles at deflector voltages of 0~kV and 30~kV are shown together with the corresponding Monte-Carlo trajectory simulations. When the deflector was turned on, the molecular beam was broadened and clearly shifted towards higher deflection coordinates due to the different deflection of different rotational states of OCS.\\
The spatial separation of different rotational states of OCS was previously demonstrated.\cite{nielsen11a, karamatskos19a} The current setup did not allow for a state-selective detection, but the degree of deflection of individual rotational states $\ket{JM}$ of OCS could be inferred from a comparison of the experimental deflection profiles to the ones obtained from trajectory simulations. Here, $J$ and $M$ stand for the quantum numbers of the rotational angular momentum and its projection on the electric-field axis, respectively. Assuming thermal populations in the expansion, a rotational temperature T$_\text{rot}=0.55(7)$~K was determined for the molecular beam of OCS molecules. The uncertainty quoted for the rotational temperature represents the uncertainty of the fit of the simulated thermally averaged density profiles to the experimental data. The rotational states $\ket{00}$ and $\ket{11}$ were strongly strong-field-seeking at the present electric-field strengths. Although a complete separation was only achieved for the rovibrational ground state $\ket{00}$ at high deflection coordinates (Fig.~\ref{fig:Figure3}), a unique distribution of rotational-state populations was established in the beam at each deflection coordinate. This enables the determination of state-specific reaction rates from measurements of state-averaged reaction rates at different deflection coordinates.\cite{kilaj20a}

\subsection{\label{sec:level10} State-selective Chemi-ionisation Reactions of Ne\textsuperscript{*} with OCS ($J$)}
\begin{figure}[!t]
 \centering
 \includegraphics[width=\columnwidth]{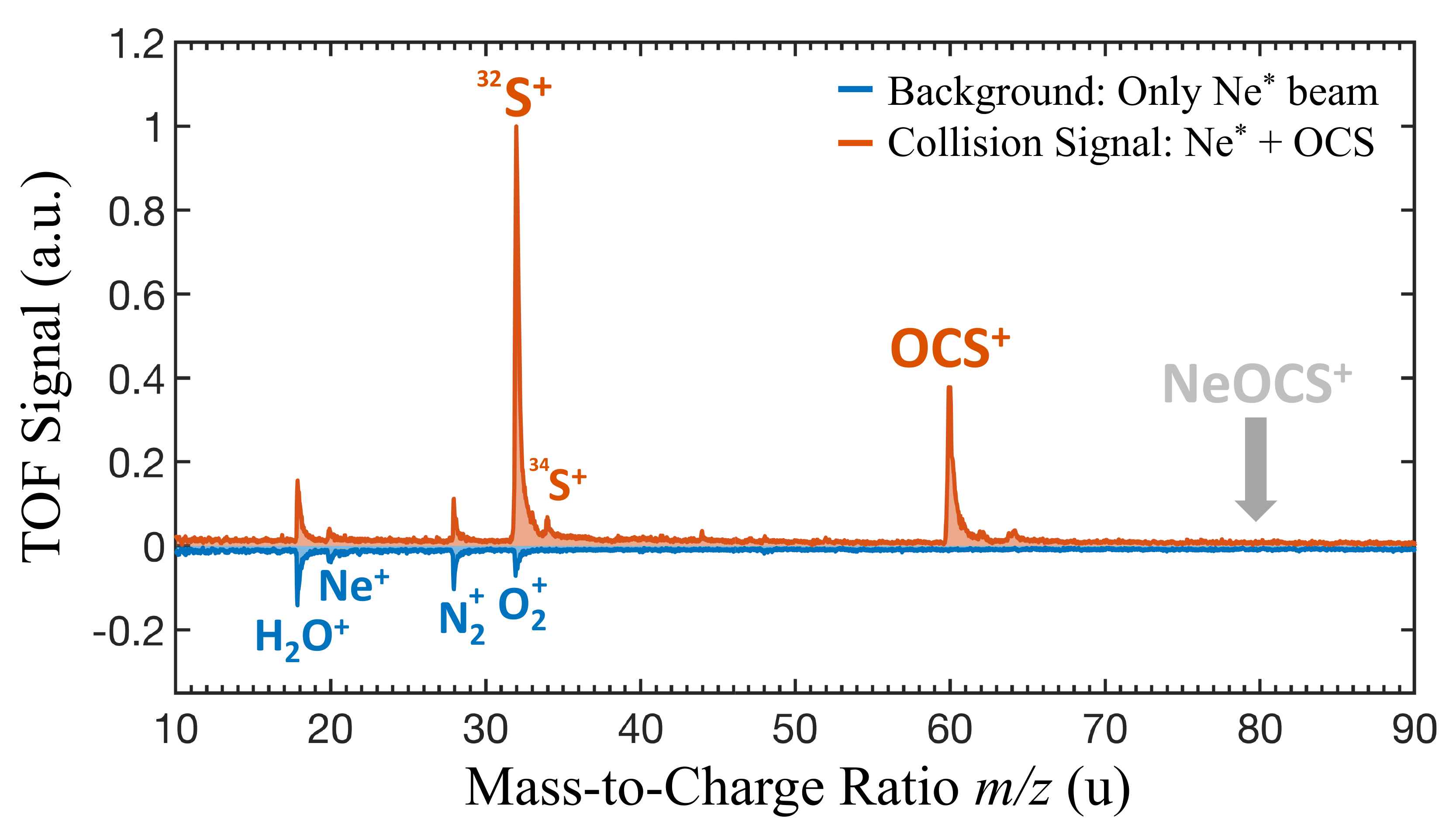}
 \caption{Time-of-flight mass spectrum of products of the chemi-ionisation reaction Ne* + OCS. Intensities are normalised with respect to the strongest signal. The trace corresponding to the background experiment without the OCS beam is inverted for clarity. The signals of the ionic products originating from Penning- and dissociative-ionisation reaction pathways (OCS$^+$ and S$^+$, respectively) are indicated. The associative-ionisation product NeOCS$^+$ is not observed within the sensitivity limits of the present experiment.}
 \label{fig:Figure4}
 \end{figure}

\begin{figure}[!t]
 \centering
 \includegraphics[width=\columnwidth]{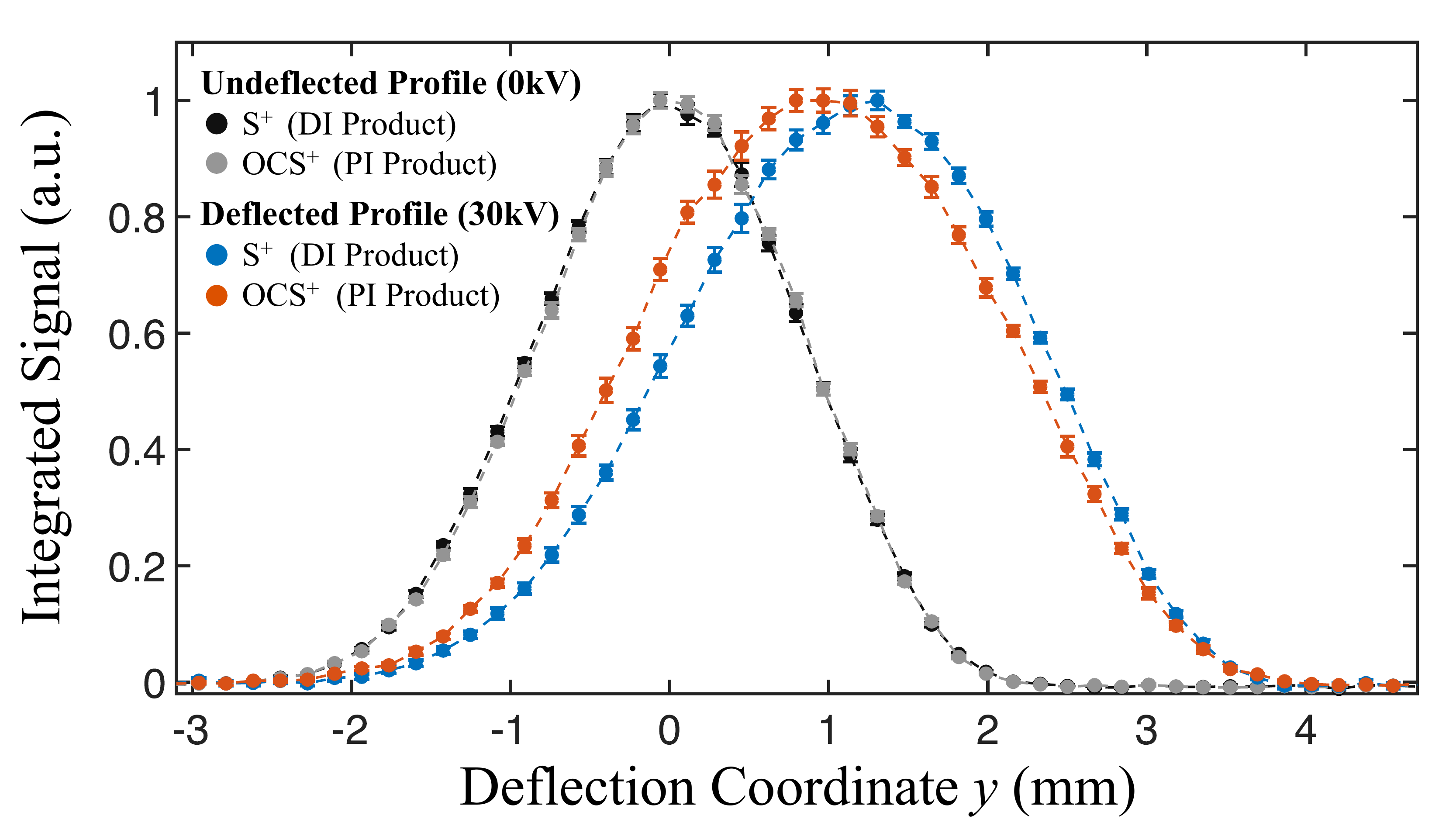}
 \caption{Normalised chemi-ionisation reaction-deflection profiles of OCS + Ne* recorded for the Penning-ionisation (PI) and dissociative-ionisation (DI) products obtained with deflected and undeflected beams. The error bars represent the standard error of 15 individual traces in which each data point was averaged over 1000 experimental cycles. In contrast to the experiment with undeflected beams (black and grey data points), a clear shift of the profiles associated with the PI and DI products is visible when the OCS molecules in different rotational states are spatially separated by the deflector (red and blue data points). The different shifts of the PI and DI product profiles are attributed to different reactivities of individual rotational states in the two reaction pathways.}
 \label{fig:Figure5} 
 \end{figure}

\begin{figure}[!t]
 \centering
 \includegraphics[width=\columnwidth]{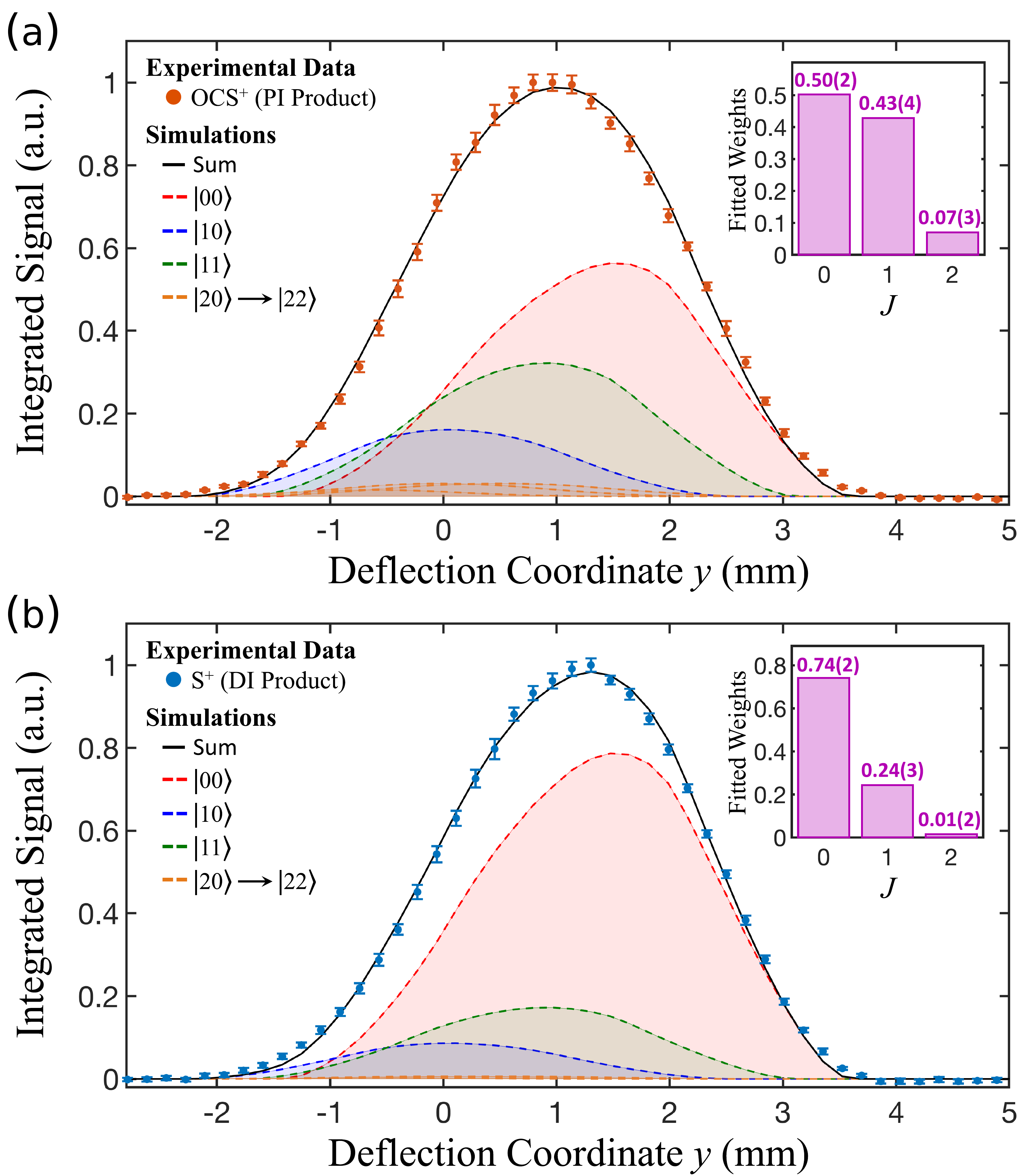}
 \caption{Least-square fits of state-specific simulated reaction-deflection profiles to the experimental data for (a) the Penning- and (b) the dissociative-ionisation products. Only the rotational states $J=0,1,2$ significantly populated in the molecular beam were taken into account (see inset in Fig.~\ref{fig:Figure3}). Insets: Relative fitted weights of the individual rotational states ($J$) of OCS contributing to the reaction-deflection profiles. The uncertainties represent the fit errors.}
 \label{fig:Figure6}
 \end{figure}

To demonstrate the capabilities of the new CMB setup, state-specific chemi-ionisation reactions of metastable neon atoms with OCS ($J$) were studied. \\
Investigations of chemi- (or Penning-) ionisation reactions have a long history.\cite{siska93a} Recently, remarkable progress has been achieved in their characterisation at very low collision energies.\cite{henson12a,falcinelli20a, dulitz20a, gordon20a,paliwal21a}
The reaction of OCS with metastable rare gas atoms has been investigated before using Penning-ionisation electron spectroscopy,\cite{cermak76a, brion77a} photoemission spectroscopy\cite{setser73a, tsuji80a} and CMB experiments which revealed strong anisotropies in the interaction potentials.\cite{kishimoto03a, horio06a} In general, the reaction Ne\textsuperscript{*} + OCS can proceed via different pathways yielding different reaction products: the Penning-ionisation (PI) products OCS\textsuperscript{+} and Ne, the associative-ionisation (AI) product (NeOCS\textsuperscript{+}) or dissociative-ionisation (DI) products of OCS. In the latter case, multiple products are possible depending on the amount of energy available in the reaction.\cite{tsuji80a} \\
In order to investigate state-specific effects in this reaction, a deflected beam of OCS molecules was intersected with a beam of metastable neon atoms. Metastable neon atoms in the (2p$^5$3s)~\textsuperscript{3}P\textsubscript{2} and \textsuperscript{3}P\textsubscript{0} states were generated in a supersonic expansion of neon gas (stagnation pressure 25~bar) through a pulsed plate-discharge source. Product ions generated by chemi-ionisation in the collision region were accelerated towards the detector by pulsing the extraction fields for 1~$\mu$s. A typical TOF mass spectrum of the reaction products is displayed in Fig.~\ref{fig:Figure4}. 
Besides products resulting from the Penning ionisation of trace gases in the background vacuum (H\textsubscript{2}O\textsuperscript{+}, N\textsubscript{2}\textsuperscript{+}, O\textsubscript{2}\textsuperscript{+}), the products of the PI pathway (OCS\textsuperscript{+}) and the dissociation product S\textsuperscript{+} are the dominating species observed in the mass spectrum. Other ionic products from the dissociation pathway are energetically unavailable according to Ref.~\onlinecite{tsuji80a} considering the excitation energy of metastable neon (16.6~eV for Ne(\textsuperscript{3}P\textsubscript{2}) and 16.7~eV for Ne~(~\textsuperscript{3}P\textsubscript{0})) and the present collision energy of 0.3~eV. From the integral of the relevant signals in the TOF mass spectrum, a total branching ratio between the PI and DI products (PI:DI) of 1:2.5 could be determined. An AI product at $m/z=$80~u was not observed, but it cannot be ruled out that part of the signal at $m/z=$60~u is due to dissociation of a weakly bound AI complex.
However, for the present collision energies of 0.3~eV we expect the contribution of the AI pathway to be negligible. The small Ne\textsuperscript{+} signal in the spectrum is attributed to intra-beam Ne\textsuperscript{*}-Ne\textsuperscript{*} PI which is also present in the background experiment with only the Ne\textsuperscript{*} beam (blue inverted trace in Fig.~\ref{fig:Figure4}). Moreover, the Ne\textsuperscript{+} signal did not decrease when the Ne beam was overlapped with the OCS expansion suggesting that possible ion-molecule reactions Ne\textsuperscript{+}~+~OCS do not play a major role under the present conditions. \\
The chemical reactivity of the different rotational states of OCS was probed by overlapping different parts of the deflected OCS beam with the Ne\textsuperscript{*} expansion in the collision chamber. TOF mass spectra of ionic reaction products were recorded for different tilting angles of the OCS molecular beam setup while keeping the position of the Ne\textsuperscript{*} beam fixed. Reaction-deflection profiles for both the PI and DI products were obtained by plotting the integrated TOF signals of the products as a function of deflection coordinate $y$. These profiles reflect the chemical reactivity of OCS in the distribution of rotational states varying across the deflected beam. Normalised profiles for the reaction products from the PI and DI channels obtained with both undeflected and deflected molecular beams are displayed in Fig.~\ref{fig:Figure5}. The profile of the DI product was corrected for a small background signal at the same mass-to-charge ratio $m/z=32$~u originating from PI of trace molecular oxygen in the background vacuum which does not vary as a function of the deflection coordinate. \\
The undeflected profiles of the PI and DI products, in which the individual rotational states of OCS were not spatially separated, overlap. By contrast, a clear shift between the PI and DI product was observed for the reaction  profiles corresponding to the deflected beams. We attribute these differences to different reaction probabilities of individual rotational states in the two reaction pathways. We assume that the OCS molecules are isotropically oriented in the scattering region, as any orientation imparted by the field of the deflector should dephase during the transit over 45~cm from the exit of the deflector assembly.\cite{kienitz16a}  \\   
The contribution of each rotational state to the PI and DI reaction-deflection profiles was determined by fitting a weighted sum of state-specific simulated profiles to the experimental data. The results are shown in Fig.~\ref{fig:Figure6}. In these fits, the properties of the molecular beam were fixed to the values obtained from fits of simulations to the fs-laser deflection profiles (see Sec. \ref{sec:level9}) and only the weights of the individual rotational profiles in the sum was adjusted. As no explicit orientation effects are expected, all $M$ states associated with a specific rotational state $J$ were weighted with a degeneracy factor $g_M=1$ for $M=0$ and $g_M=2$ for $M>0$ in the fit. 
Furthermore, as the beam waist of the fs-laser focus ($w\approx30~\mu$m) is much smaller than the width of the second molecular beam ($d\approx1.5$~mm), the different sampling volumes in the fs-laser ionisation and reaction experiments were taken into account in the simulations. \\
As can be seen from the relative fitted weights of the different rotational states to the PI and DI reaction-deflection profiles in the insets of Figs.~\ref{fig:Figure6} (a) and (b), the individual states contribute with different efficiencies to the PI and DI pathways. OCS molecules in the rotational ground state $J=0$ are about a factor 2.5 more reactive in dissociative ionisation (yielding S\textsuperscript{+}) ions than in Penning ionisation (producing OCS\textsuperscript{+}) in comparison to the $J=1$ state. A detailed analysis of these findings will be published elsewhere. These results demonstrate the utility of the present setup to isolate individual rotational states of a molecule and study their reactivity in a crossed-beam experiment.

\section{\label{sec:level11}Conclusion and Outlook \protect}
We have presented a novel CMB experiment for the investigation of state- and conformer-specific reactions and demonstrated its capabilities for studying rotational-state effects in Ne\textsuperscript{*} + OCS chemi-ionistion. We found that the branching ratio between the Penning and dissociative ionisation pathways strongly depends on the initial rotational state of the OCS molecule. The present experiment introduces a new approach to study rotational effects in chemi-ionisation and related reactions. \\
Besides the investigation of state-specific reactions demonstrated here, the novel CMB setup is also suited for studies of conformational effects in reactions of complex molecules under single-collision conditions, in analogy to our previous studies on conformationally selected ion-molecule reactions.\cite{chang13a, roesch14a} As conformers of a wide range of molecules have already been spatially separated with the present method,\cite{filsinger09a, chang13a, roesch14a, wang20, teschmit18a, you18a} a variety of reactions is amenable to the present approach. The ability of the electrostatic deflector to select different sizes of molecular clusters\cite{trippel12a, chang15a, you18a} would allow for reactions with even more complex reactants.  \\ 
Moreover, the discharge source developed for the present setup is also capable of producing high-density molecular beams of free radicals\cite{stranak21a} paving the way for the investigation of state- and conformer-specific effects in reactions with these species in the present setup. \\

\begin{acknowledgments}
We acknowledge Philipp Kn\"opfel, Grischa Martin, Georg Holderried and Anatoly Johnson for technical support. We also would like to thank Kopin Liu for valuable suggestions and fruitful discussions during the conception of the experimental setup. We thank Prof. Frank Stienkemeier (Freiburg) for providing us with a CRUCS valve used in the present experiments. This work is supported by Swiss National Science Foundation under grant nr. BSCGI0\_15787 and the University of Basel. H. G. and S. W. acknowledge support by the K. C. Wong Education Foundation.
\end{acknowledgments}


\end{document}